%% file: main.tex
\documentclass[nonacm, language=english, anonymous=false]{acmart}
\usepackage{hyperref}
\usepackage[utf8]{inputenc}
\usepackage{url}

\usepackage{breakurl}
\usepackage{wrapfig}
\usepackage{graphicx}
\usepackage{subcaption}
\usepackage{float}
\usepackage{multirow}
\usepackage{amsmath}

\begin{document}
\title{Quantifying the Risk of Transferred Black Box Attacks}

\author{Disesdi Susanna Cox}
\authornote{Both authors contributed equally to the paper}
\orcid{0009-0003-0568-0236}
\affiliation{%
	\institution{OWASP AI Exchange}
	\streetaddress{300 Delaware Ave, Ste 210 \#384}
	\city{Wilmington}
	\state{DE}
	\country{USA}
	\postcode{19801}
}
\email{disesdi.susannacox@owasp.org}

\author{Niklas Bunzel}
\authornotemark[1]
\orcid{0000-0002-8921-1562}
\affiliation{%
	\institution{Fraunhofer SIT / TU-Darmstadt / ATHENE}
	\streetaddress{Rheinstraße 75}
	\city{Darmstadt}
	\state{Hesse}
	\country{Germany}
	\postcode{64295}
}
\email{niklas.bunzel@sit.fraunhofer.de}

\begin{abstract}
Neural networks have become pervasive across various applications, including security-related products. However, their widespread adoption has heightened concerns regarding vulnerability to adversarial attacks. With emerging regulations and standards emphasizing security, organizations must reliably quantify risks associated with these attacks, particularly regarding transferred adversarial attacks, which remain challenging to evaluate accurately.
This paper investigates the complexities involved in resilience testing against transferred adversarial attacks. Our analysis specifically addresses black-box evasion attacks, highlighting transfer-based attacks due to their practical significance and typically high transferability between neural network models. We underline the computational infeasibility of exhaustively exploring high-dimensional input spaces to achieve complete test coverage. As a result, comprehensive adversarial risk mapping is deemed impractical.
To mitigate this limitation, we propose a targeted resilience testing framework that employs surrogate models strategically selected based on Centered Kernel Alignment (CKA) similarity. By leveraging surrogate models exhibiting both high and low CKA similarities relative to the target model, the proposed approach seeks to optimize coverage of adversarial subspaces. Risk estimation is conducted using regression-based estimators, providing organizations with realistic and actionable risk quantification.
\end{abstract}



\maketitle

\section{Introduction}
\input{content/intro}
\section{Related Work}
\input{content/sota}
\section{Neural Network Similarity: CKA}
\input{content/background}
\section{Coverage Testing}
\input{content/coverage}
\section{Conclusion \& Future Work}
\input{content/conclusion}

\begin{acks}
This research work has been funded by the German Federal Ministry of Education and Research and the Hessian Ministry of Higher Education, Research, Science and the Arts within their joint support of the National Research Center for Applied Cybersecurity ATHENE.
\end{acks}

\bibliographystyle{ACM-Reference-Format}
\bibliography{mybib}

\end{document}

%% file: content/intro.tex
Neural networks have become widespread in numerous commercial applications, including critical security-related products. Despite their impressive capabilities, neural networks remain highly vulnerable to evasion attacks, wherein attackers subtly manipulate inputs to alter model predictions in their favor.
With increasing recognition of these vulnerabilities, political and legislative frameworks worldwide are rapidly evolving to mandate the secure deployment of AI systems. Prominent among these is the EU AI Act, which represents one of the pioneering comprehensive regulations requiring organizations to ensure AI robustness and security. Concurrently, international standards for AI security are emerging, compelling companies to demonstrate compliance with established security criteria.
In the realm of evasion attacks, current research has proposed various defensive strategies, including adversarial training~\cite{pgd_bib}, preprocessing techniques such as denoising~\cite{9010982} and JPEG compression~\cite{das2017keeping}, and detectors based on image statistics~\cite{bunzel2024signals} or neuron activations~\cite{roth2019odds}. However, regulatory compliance now requires organizations not merely to defend but also to quantify and demonstrate the associated adversarial risk explicitly.
As regulatory bodies and other agencies tasked with ensuring AI system security grapple with setting normative standards and regulatory thresholds, successfully integrating risk assessment into AI security workflows is becoming increasingly essential. Organizations require means of quantifying risk and resilience to various AI security vectors for both regulatory compliance, and overall system security.
Recent literature presents several risk estimation methodologies, predominantly relying on red teaming approaches, where trained models are subjected to various optimization-based attacks~\cite{croce2020robustbench, Croce020a, guo2024exploringadversarialfrontierquantifying}. However, these methodologies exhibit gaps in assessing risks associated with: 
\begin{itemize}
    \item Transferred adversarial attacks, which are particularly effective in real-world scenarios
    \item Attacks transferred from an extensive range of potential surrogate models
    \item Simultaneous consideration of multiple attack types
    \item Performing risk estimations earlier in the model lifecycle e.g. pre-training or during design phase
\end{itemize}

Addressing these critical gaps, this paper introduces a comprehensive framework designed explicitly for estimating risks associated with transferred adversarial attacks. A central contribution is our systematic and extensive coverage of surrogate model spaces, ensuring robust risk quantification. This enables organizations to effectively assess adversarial vulnerabilities.

%% file: content/sota.tex
\subsection{Adversarial Attacks}
Adversarial attacks manipulate neural network outputs by introducing small, carefully crafted perturbations to input data. Szegedy et al.~\cite{fgsm_bib} demonstrated that minimal, often imperceptible, perturbations could drastically alter deep neural network predictions. Initially observed in image classification, this vulnerability extends to semantic segmentation~\cite{arnab2018robustness}, object detection~\cite{MI2023114, yin2021adc}, tracking~\cite{SUTTAPAK202221}, natural language processing~\cite{zhang2019adversarial}, and speech recognition~\cite{zelasko2021adversarial}. Adversarial perturbations are particularly critical as they induce high-confidence misclassifications and generalize across models~\cite{transferability_0}. While input-specific perturbations target individual samples, universal perturbations effectively mislead multiple inputs and models~\cite{moosavidezfooli2017universal}.

\paragraph{Transferability}
Transferability refers to the ability of adversarial examples crafted for one model to mislead different models with varying architectures and training data.
First observed in~\cite{szegedyIntriguingPropertiesNeural2014}, transferability is linked to adversarial perturbations aligning with model weight vectors and complexity~\cite{DemontisMPJBONR19, klause2025relationshipnetworksimilaritytransferability}.
Petrov et al.~\cite{transferability_1} and Alvarez et al.~\cite{alvarezExploringTransferabilityAdversarial2023} found perturbation similarities across related architectures.

\subsection{Risk Estimation}
Previous research has primarily investigated adversarial risk by measuring model accuracy under adversarially perturbed inputs~\cite{fgsm_bib,pgd_bib}. While these individual evaluations provide useful insights, they fail to capture the full spectrum of adversarial risk, as a model's robustness can vary significantly depending on the type of attack and the perturbation budget. 
RobustBench~\cite{croce2020robustbench}, a standardized benchmark for adversarial robustness, addresses this issue by incorporating AutoAttack~\cite{Croce020a}, an ensemble of diverse and adaptive adversarial attacks. AutoAttack combines several attack strategies, including APGD~\cite{Croce020a}, FAB~\cite{Croce020}, and Square Attack~\cite{squareattack_bib}, to provide a more comprehensive robustness assessment. By evaluating models against a curated set of strong attacks, RobustBench helps mitigate the limitations of single-attack evaluations.
However, adversarial robustness depends not only on attack diversity, but also on the perturbation budget, which defines the maximum allowable change to an input. Evaluating robustness over different perturbation budgets is essential to understanding the behavior of the model under different attack strengths. For example,~\cite{guo2024exploringadversarialfrontierquantifying} shows that robustness metrics can shift significantly when perturbation constraints are relaxed or tightened. Without such multi-scale evaluations, adversarial risk assessments remain incomplete, potentially leading to misleading conclusions about a model's true robustness.
In~\cite{TramerB19} the authors highlight the trade-offs in multi-perturbation robustness and provide insights into how different perturbation types affect adversarial risk.
Given the diversity of attack strategies, the current state of the art covers query-based white-box and black-box attacks, including gradient-based methods and optimization-based attacks. Risk estimation for transfer-based approaches is investigated in~\cite{klause2025relationshipnetworksimilaritytransferability} on a per model basis.

%% file: content/background.tex
Several methods exist for evaluating similarity of neural networks. Canonical Correlation Analysis (CCA)~\cite{cca_bib} is a statistical method for finding relationships among variables which has been adapted to use in comparison of activation patterns in neural networks. Similarly, Singular Vector Canonical Correlation Analysis (SVCCA)~\cite{svcca_bib} augments the CCA methodology by forming the comparison after a dimensionality reduction of the activation space via singular value decomposition (SVD). This dimensionality reduction results in increased robustness to noise, and greater utility in high-dimensional representations. Both CCA and SVCCA rely on linear correlation.

In contrast, Centered Kernel Alignment (CKA)\cite{cka_bib} can be used to evaluate similarities among networks with complex, non-linear transformations, and is thus useful in application to both linear and non-linear representations. CKA utilizes a normalization of the Hilbert-Schmidt Independence Criterion (HSIC)~\cite{GrettonBSS05} as a means to non-parametrically measure variable independence, comparing similarity matrices of the activations induced by a set of inputs across two or more models. This results in a similarity score between 0 and 1. By evaluating the similarities in how various networks respond to identical inputs, CKA is a useful tool for understanding how models map relationships, and how these learned mappings differ among models. A CKA network similarity score provides an estimation of similarity between networks. Due to its robustness in capturing representational similarity, CKA is the chosen method for comparing CNNs in this work.

%% file: content/coverage.tex
\begin{figure*}[ht!]
    \centering
    \includegraphics[width=\linewidth]{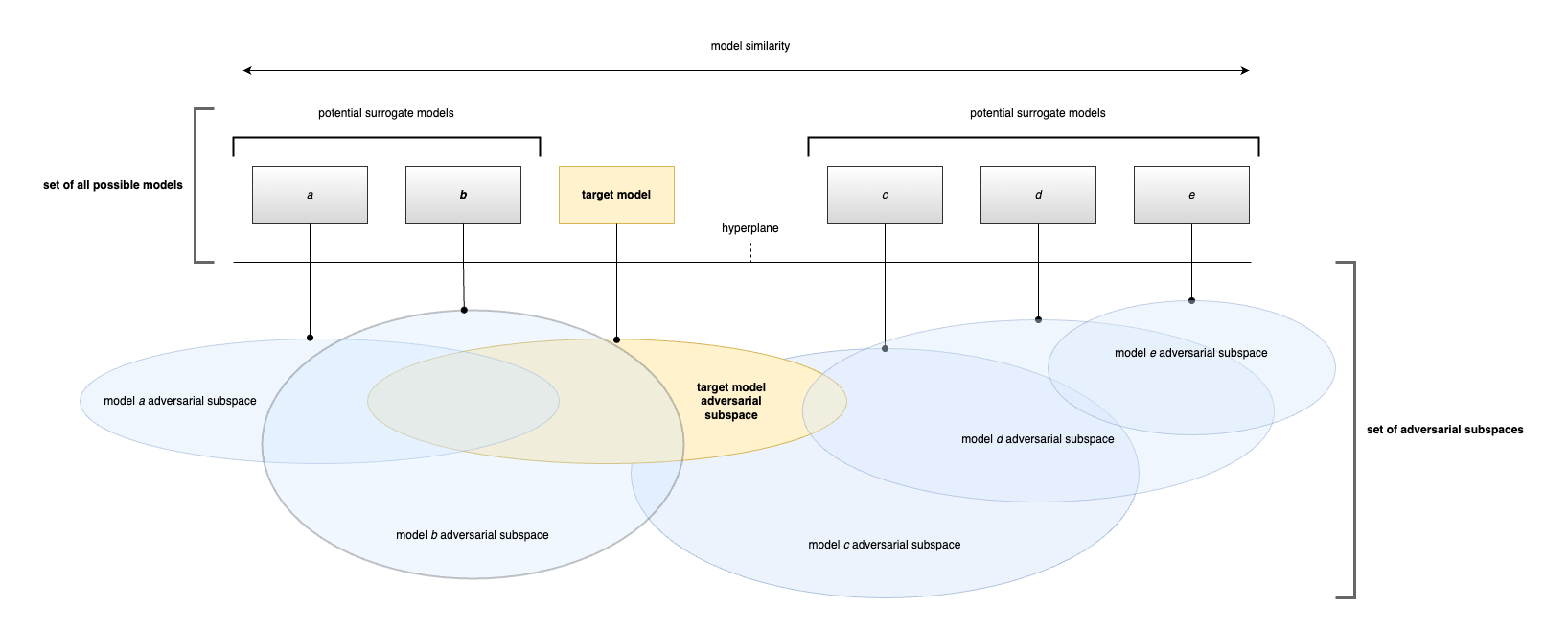}
    \caption{Conceptual illustration of adversarial subspace overlap across models with varying similarity. Target model (yellow) and potential surrogate models (a–e), each with its own adversarial subspace.}
    \label{fig:theory}
\end{figure*}

\subsection{Full-Coverage Testing Feasibility: Transferability, Scope, and Search Space}
Black-box poisoning attacks present a range of most-likely attack scenarios; it is assumed that attackers have partial but incomplete knowledge of systems and data, versus white-box attacks, in which the adversary has total access to all critical system parameters and assets. Black-box attacks thus likely mirror real-life scenarios more closely than white-box methods. Black-box attacks may be classified as transfer- or query-based. Query-based methods require an ability on the attacker’s part to probe a system and iteratively refine attacks, whereas transfer-based attacks attempt to construct an approximate surrogate model, and design optimized attacks against the surrogate which are intended to transfer to the target model~\cite{CinaGDVZMOBPR23}. A number of attack methods found in the literature are transfer-based, and attack transferability is found to be generally high~\cite{DemontisMPJBONR19}. Attack transferability thus plays a significant role in data poisoning.\\
Goodfellow et al (2014) noted that adversarial samples occur in large, contiguous subspaces, rather than randomly scattered pockets~\cite{fgsm_bib}. Tramèr et al (2017) found that transferable adversarial examples may span a contiguous subspace of $\sim 25$ dimensionality, which is large. The dimensionality of these subspaces is relevant to transferability because the higher the dimensionality, the more likely it is that the subspaces of two models will intersect significantly, thus enabling transfer~\cite{transferability_3}.\\
Attacks are largely model agnostic, with transferability enabled by similarities in decision boundaries which are shared by models of different classes, and may not be displaced by adversarial training~\cite{transferability_3}. Given the large adversarial sample space, as well as the nature of transferability vis-a-vis decision boundary similarity, defining a subset of applicable attacks and corresponding tests for a given model system may be infeasible.\\
If defining appropriate testing for full coverage requires finding all intersecting adversarial subspaces, achieving full testing coverage may be infeasible.\\
Demontis et al (2018) give a formal definition for transferability that depends on: (1) the size of input gradients of the target classifier; (2) how well the gradients of the surrogate and target models align; and (3) the variance of the loss landscape optimized to generate the attack points [4]. Given the high importance of transferability in poisoning scenarios, the number of potential applicable attacks (and corresponding tests) would likely correlate to these three metrics, evaluated on a per-model basis. This presents a challenging combinatorial analysis problem, in addition to the difficulty of defining all intersecting subspaces.\\
Ding et al (2020) formulate robust optimization against outliers as a combinatorial optimization problem, showing that even the simplest one-class SVM with outliers problem is NP-complete, with no fully polynomial-time approximation scheme unless P = NP, meaning that it is unlikely to achieve a near-optimal solution in polynomial time~\cite{Ding20}. This further suggests infeasibility of anticipating and testing for every potential vector.\\
Conclusion: both empirical data and formal methods cast doubt on the feasibility of achieving full or even adequate coverage via resilience testing.

\subsection{Best-Coverage Testing: Exploiting Adversarial Subspaces to Increase Coverage and Attackers' Costs}
While we cannot cover all adversarial subspaces, we can try to maximize our coverage \& increase attacker costs. In order to increase testing coverage, we may begin by deriving some n number of surrogate models which do not have overlapping adversarial subspaces, and test using these. If the subspaces overlap, there may be reduced value to testing with more than one surrogate. Determining overlap is key, yet mapping subspaces is likely infeasible.\\
Since we cannot easily map the subspaces, we may instead choose surrogates which exhibit both high and low similarity to the model being tested. In absence of the ability to prove the lack of subspace overlap, we may minimize the probability of testing on overlapping subspaces by choosing both quantitatively similar and different surrogates to test. The identification of these surrogates first requires selection of a metric by which to estimate their similarity.
The concept is illustrated in Figure~\ref{fig:theory}, the adversarial subspace overlap across models with varying similarity. The target model is surrounded by potential surrogate models, each model induces its own adversarial subspace, with the degree of intersection indicating the potential for transferability of adversarial examples. Surrogate models more similar to the target model (e.g., models a and b) exhibit greater overlap with the target model’s adversarial subspace, suggesting higher attack transferability. In contrast, dissimilar models (e.g., d and e) show reduced overlap, limiting the effectiveness of transferred attacks.

\subsection{Similarity Thresholds Via Centered Kernel Alignment}
Achieving maximum adversarial subspace coverage, with the goal of increasing cost of attacks, likely requires diversity of test surrogates. However, defining the number and metric for diversity of surrogates is challenging. How many surrogates we choose would ideally be tied to complexity of the pre-production model in testing; however this is difficult to quantify currently and thus to scale as a requirement for production systems. Ideally tests could be performed to determine subspace overlap (in order to avoid overlap and achieve more coverage), but this is currently infeasible. In this light, a means of comparing similarity of networks may serve as a proxy for evaluating potential subspace overlap.

Previous work has demonstrated effectiveness of CKA similarity scores for estimation of black-box attack success rates~\cite{klause2025relationshipnetworksimilaritytransferability}. We extend this work here to provide a methodology for risk estimation; specifically, quantification of the risk of transferability of black-box attacks.

\subsection{A More Formal Definition}
Since evaluating transferability of all potential attacks effective against all potential surrogate models is infeasible, we may define a subset of surrogate models via their measured similarity to one another, and test against these. To ensure maximum coverage of subspaces, we apply our similarity definition to the selection of both highly similar, and highly disparate, models.

In order to achieve testing coverage of as disparate a set of adversarial subspaces as reasonably feasible, testing should consist of surrogate models calculated to have both high and low CKA similarity to the pre-production model in testing $M_{p}$. For convenience, we may discuss these loosely as subsets of the whole set of surrogate models whose derivations might exist in shared adversarial subspaces. We can thus define some $n$ number of requisite surrogate models to be derived and tested against the pre-production model for attack transferability. Due to the high importance of transferability to adversarial attack susceptibility, and the relationships among model similarity and complexity to the likelihood of extant shared adversarial subspaces, organizations should test using $n > 2$ derived surrogate models $S$, with at least $n \geq 1$ surrogate models from a pool of highly-similar networks, and $n \geq 1$ surrogate models from a pool of low-similarity networks, quantified on the basis of their calculated CKA similarity to the pre-production model in testing. Increasing $n$ (e.g., $n \geq 5$) provides greater statistical significance and may lead to better coverage of diverse adversarial spaces.

Let $M_{1}$ represent the members of a subset of surrogate models exhibiting close CKA similarity to $M_{p}$. Let $M_{2}$ represent the members of a subset of surrogate models found to have lower CKA similarity relative to $M_{p}$. Let $|M_{1}|$ and $|M_{2}|$ represent the cardinality of $M_{1}$ and $M_{2}$, respectively.

Let $r$ represent the CKA similarity threshold for each model subset. Let $r_{1}$ represent the lower bound of CKA similarity between the pre-production model and surrogate(s) in $M_{1}$. Let $r_{2}$ represent the upper bound of CKA similarity between the pre-production model and members of $M_{2}$.

We can thus define:

$M_{1} = \{S_i \mid \text{CKA}(S_i, M_p) \geq r_{1}\}$, and 
$M_{2} = \{S_j \mid \text{CKA}(S_j, M_p) \leq r_{2}\}$ with $r_{1}$ and $r_{2}$ satisfying $0 < r_{2} < r_{1} < 1$.

To maximize potential coverage, organizations should perform pre-release resilience testing consisting of attacks derived from some

$|M_{1}| \geq 1$, and 
$|M_{2}| \geq 1$ 

number of surrogate models from their respective complexity threshold groups.

\paragraph{Selecting the Thresholds}
Concrete guidance for defining suitable threshold values for $r_{1}$ and $r_{2}$ can be derived empirically. For instance, selecting $r_{1} \approx 0.55$ and $r_{2} \approx 0.35$ may be appropriate, given that most CNN architectures exhibit a CKA similarity score between 0.32 and 0.57, with a median of approximately 0.45~\cite{klause2025relationshipnetworksimilaritytransferability}. However, the precise determination of these thresholds should be contextually informed by initial empirical experimentation and considerations specific to the domain of application.

Building upon the CKA metric, the Diagonal Box Similarity (DBS) score was introduced in~\cite{klause2025relationshipnetworksimilaritytransferability} to more effectively quantify localized layer-to-layer similarity between neural networks. DBS scores for most CNN architectures typically range from 0.4 to 0.75, thereby facilitating more granular and precise threshold selections, such as $r_{1}\approx 0.7$ and $r_{2}\approx 0.45$.

%% file: content/conclusion.tex
This paper addresses the challenges in resilience testing against transferred black-box adversarial evasion attacks, emphasizing transfer-based methods due to their practical relevance and inherent high transferability among neural network models. Recognizing the computational infeasibility of exhaustive exploration of high-dimensional input spaces, the paper argues against attempting complete test coverage for adversarial risk mapping.
To overcome these constraints, we introduce a targeted resilience testing framework utilizing surrogate models strategically selected via Centered Kernel Alignment (CKA) similarity. By incorporating surrogate models with both high and low CKA similarity relative to the target model, the proposed framework enhances coverage of adversarial subspaces. Risk estimation within this approach employs regression-based estimators, offering practical, realistic quantification of adversarial risk.
Furthermore, the framework supports model selection strategies aimed at minimizing susceptibility, thus elevating attacker costs and reducing exposure. The approach incorporates clear, threshold-based criteria to facilitate straightforward implementation in resilience testing processes.
Given the increasing regulatory emphasis on integrating quantitative risk assessment into AI security standards, the proposed methodology provides an efficient and actionable solution. It enables organizations to achieve accurate risk quantification and regulatory compliance.
Future work will focus on developing more accurate similarity metrics, with an emphasis on improving computational efficiency. Ideally, these metrics should be derived from the model architecture itself rather than relying on activations. By leveraging structural properties and design patterns of neural networks, we aim to create methods that offer faster and more scalable comparisons while maintaining high reliability in assessing model similarity, with the ultimate goal of providing increasingly useful and cost-effective risk quantification methods for AI systems in production.

%% file: mybib.bib
@inproceedings{GrettonBSS05,
  author       = {Arthur Gretton and
                  Olivier Bousquet and
                  Alexander J. Smola and
                  Bernhard Sch{\"{o}}lkopf},
  editor       = {Sanjay Jain and
                  Hans Ulrich Simon and
                  Etsuji Tomita},
  title        = {Measuring Statistical Dependence with Hilbert-Schmidt Norms},
  booktitle    = {Algorithmic Learning Theory, 16th International Conference, {ALT}
                  2005, Singapore, October 8-11, 2005, Proceedings},
  series       = {Lecture Notes in Computer Science},
  volume       = {3734},
  pages        = {63--77},
  publisher    = {Springer},
  year         = {2005},
}

@article{CinaGDVZMOBPR23,
  author       = {Antonio Emanuele Cin{\`{a}} and
                  Kathrin Grosse and
                  Ambra Demontis and
                  Sebastiano Vascon and
                  Werner Zellinger and
                  Bernhard Alois Moser and
                  Alina Oprea and
                  Battista Biggio and
                  Marcello Pelillo and
                  Fabio Roli},
  title        = {Wild Patterns Reloaded: {A} Survey of Machine Learning Security against
                  Training Data Poisoning},
  journal      = {{ACM} Comput. Surv.},
  volume       = {55},
  number       = {13s},
  pages        = {294:1--294:39},
  year         = {2023},
  url          = {https://doi.org/10.1145/3585385},
  doi          = {10.1145/3585385}
}

@misc{Ding20,
  author       = {Hu Ding and
                  Fan Yang and
                  Jiawei Huang},
  title        = {Defending Support Vector Machines against Poisoning Attacks: the Hardness
                  and Algorithm},
  year         = {2020},
  url          = {https://arxiv.org/abs/2006.07757},
  eprinttype    = {arXiv},
  eprint       = {2006.07757}
}

@misc{klause2025relationshipnetworksimilaritytransferability,
      title={The Relationship Between Network Similarity and Transferability of Adversarial Attacks}, 
      author={Gerrit Klause and Niklas Bunzel},
      year={2025},
      eprint={2501.18629},
      archivePrefix={arXiv},
      primaryClass={cs.CR},
      url={https://arxiv.org/abs/2501.18629}, 
}

@misc{fgsm_bib,
      title={Explaining and Harnessing Adversarial Examples}, 
      author={Ian J. Goodfellow and Jonathon Shlens and Christian Szegedy},
      year={2015},
      eprint={1412.6572},
      archivePrefix={arXiv},
      primaryClass={stat.ML}
}

@misc{pgd_bib,
      title={Towards Deep Learning Models Resistant to Adversarial Attacks}, 
      author={Aleksander Madry and Aleksandar Makelov and Ludwig Schmidt and Dimitris Tsipras and Adrian Vladu},
      year={2019},
      eprint={1706.06083},
      archivePrefix={arXiv},
      primaryClass={stat.ML}
}

@misc{squareattack_bib,
      title={Square Attack: a query-efficient black-box adversarial attack via random search}, 
      author={Maksym Andriushchenko and Francesco Croce and Nicolas Flammarion and Matthias Hein},
      year={2020},
      eprint={1912.00049},
      archivePrefix={arXiv},
      primaryClass={cs.LG}
}

@misc{cka_bib,
      title={Similarity of Neural Network Representations Revisited}, 
      author={Simon Kornblith and Mohammad Norouzi and Honglak Lee and Geoffrey Hinton},
      year={2019},
      eprint={1905.00414},
      archivePrefix={arXiv},
      primaryClass={cs.LG}
}

@article{cca_bib,
  author={Hardoon, David R. and Szedmak, Sandor and Shawe-Taylor, John},
  journal={Neural Computation}, 
  title={Canonical Correlation Analysis: An Overview with Application to Learning Methods}, 
  year={2004},
  volume={16},
  number={12},
  pages={2639-2664},
  doi={10.1162/0899766042321814}}

@misc{svcca_bib,
      title={SVCCA: Singular Vector Canonical Correlation Analysis for Deep Learning Dynamics and Interpretability}, 
      author={Maithra Raghu and Justin Gilmer and Jason Yosinski and Jascha Sohl-Dickstein},
      year={2017},
      eprint={1706.05806},
      archivePrefix={arXiv},
      primaryClass={stat.ML}
}

@misc{yin2021adc,
      title={ADC: Adversarial attacks against object Detection that evade Context consistency checks}, 
      author={Mingjun Yin and Shasha Li and Chengyu Song and M. Salman Asif and Amit K. Roy-Chowdhury and Srikanth V. Krishnamurthy},
      year={2021},
      eprint={2110.12321},
      archivePrefix={arXiv},
      primaryClass={cs.CV}
}

@misc{zhang2019adversarial,
      title={Adversarial Attacks on Deep Learning Models in Natural Language Processing: A Survey}, 
      author={Wei Emma Zhang and Quan Z. Sheng and Ahoud Alhazmi and Chenliang Li},
      year={2019},
      eprint={1901.06796},
      archivePrefix={arXiv},
      primaryClass={cs.CL}
}

@misc{zelasko2021adversarial,
      title={Adversarial Attacks and Defenses for Speech Recognition Systems}, 
      author={Piotr Żelasko and Sonal Joshi and Yiwen Shao and Jesus Villalba and Jan Trmal and Najim Dehak and Sanjeev Khudanpur},
      year={2021},
      eprint={2103.17122},
      archivePrefix={arXiv},
      primaryClass={eess.AS}
}

@misc{arnab2018robustness,
      title={On the Robustness of Semantic Segmentation Models to Adversarial Attacks}, 
      author={Anurag Arnab and Ondrej Miksik and Philip H. S. Torr},
      year={2018},
      eprint={1711.09856},
      archivePrefix={arXiv},
      primaryClass={cs.CV}
}

@article{MI2023114,
      title = {Adversarial examples based on object detection tasks: A survey},
      journal = {Neurocomputing},
      volume = {519},
      pages = {114-126},
      year = {2023},
      issn = {0925-2312},
      doi = {https://doi.org/10.1016/j.neucom.2022.10.046},
      url = {https://www.sciencedirect.com/science/article/pii/S0925231222013273},
      author = {Jian-Xun Mi and Xu-Dong Wang and Li-Fang Zhou and Kun Cheng},
}

@article{SUTTAPAK202221,
      title = {Diminishing-feature attack: The adversarial infiltration on visual tracking},
      journal = {Neurocomputing},
      volume = {509},
      pages = {21-33},
      year = {2022},
      issn = {0925-2312},
      doi = {https://doi.org/10.1016/j.neucom.2022.08.071},
      url = {https://www.sciencedirect.com/science/article/pii/S0925231222010633},
      author = {Wattanapong Suttapak and Jianfu Zhang and Liqing Zhang},
}

@INPROCEEDINGS{9010982,
  author={Gupta, Puneet and Rahtu, Esa},
  booktitle={2019 IEEE/CVF International Conference on Computer Vision (ICCV)}, 
  title={CIIDefence: Defeating Adversarial Attacks by Fusing Class-Specific Image Inpainting and Image Denoising}, 
  year={2019},
  volume={},
  number={},
  pages={6707-6716},
  doi={10.1109/ICCV.2019.00681}
}

@misc{transferability_0,
  title = {Delving into {{Transferable Adversarial Examples}} and {{Black-box Attacks}}},
  author = {Liu, Yanpei and Chen, Xinyun and Liu, Chang and Song, Dawn},
  year = {2017},
  number = {arXiv:1611.02770},
  eprint = {1611.02770},
  primaryclass = {cs},
  publisher = {{arXiv}},
  archiveprefix = {arxiv}
}

@misc{transferability_1,
  title = {Measuring the {{Transferability}} of {{Adversarial Examples}}},
  author = {Petrov, Deyan and Hospedales, Timothy M.},
  year = {2019},
  number = {arXiv:1907.06291},
  eprint = {1907.06291},
  primaryclass = {cs, stat},
  publisher = {{arXiv}},
  archiveprefix = {arxiv}
}

@misc{transferability_3,
  title = {The {{Space}} of {{Transferable Adversarial Examples}}},
  author = {Tram{\`e}r, Florian and Papernot, Nicolas and Goodfellow, Ian and Boneh, Dan and McDaniel, Patrick},
  year = {2017},
  number = {arXiv:1704.03453},
  eprint = {1704.03453},
  primaryclass = {cs, stat},
  publisher = {{arXiv}},
  archiveprefix = {arxiv}
}

@misc{szegedyIntriguingPropertiesNeural2014,
  title = {Intriguing Properties of Neural Networks},
  author = {Szegedy, Christian and Zaremba, Wojciech and Sutskever, Ilya and Bruna, Joan and Erhan, Dumitru and Goodfellow, Ian and Fergus, Rob},
  year = {2014},
  number = {arXiv:1312.6199},
  eprint = {1312.6199},
  primaryclass = {cs},
  publisher = {{arXiv}},
  archiveprefix = {arxiv}
}

@article{alvarezExploringTransferabilityAdversarial2023,
  title = {Exploring {{Transferability}} on {{Adversarial Attacks}}},
  author = {{\'A}lvarez, Enrique and {\'A}lvarez, Rafael and Cazorla, Miguel},
  year = {2023},
  journal = {IEEE Access},
  volume = {11},
  pages = {105545--105556},
  issn = {2169-3536},
  doi = {10.1109/ACCESS.2023.3319389}
}

@misc{moosavidezfooli2017universal,
      title={Universal adversarial perturbations}, 
      author={Seyed-Mohsen Moosavi-Dezfooli and Alhussein Fawzi and Omar Fawzi and Pascal Frossard},
      year={2017},
      eprint={1610.08401},
      archivePrefix={arXiv},
      primaryClass={cs.CV}
}

@inproceedings{DemontisMPJBONR19,
  author       = {Ambra Demontis and
                  Marco Melis and
                  Maura Pintor and
                  Matthew Jagielski and
                  Battista Biggio and
                  Alina Oprea and
                  Cristina Nita{-}Rotaru and
                  Fabio Roli},
  editor       = {Nadia Heninger and
                  Patrick Traynor},
  title        = {Why Do Adversarial Attacks Transfer? Explaining Transferability of
                  Evasion and Poisoning Attacks},
  booktitle    = {28th {USENIX} Security Symposium, {USENIX} Security 2019, Santa Clara,
                  CA, USA, August 14-16, 2019},
  pages        = {321--338},
  publisher    = {{USENIX} Association},
  year         = {2019},
  url          = {https://www.usenix.org/conference/usenixsecurity19/presentation/demontis}
}

@inproceedings{bunzel2024signals,
  title={Signals Are All You Need: Detecting and Mitigating Digital and Real-World Adversarial Patches Using Signal-Based Features},
  author={Bunzel, Niklas and Frick, Raphael Antonius and Klause, Gerrit and Schwarte, Aino and Honermann, Jonas},
  booktitle={Proceedings of the 2nd ACM Workshop on Secure and Trustworthy Deep Learning Systems},
  pages={24--34},
  year={2024}
}

@inproceedings{roth2019odds,
  title={The odds are odd: A statistical test for detecting adversarial examples},
  author={Roth, Kevin and Kilcher, Yannic and Hofmann, Thomas},
  booktitle={International Conference on Machine Learning},
  pages={5498--5507},
  year={2019},
  organization={PMLR}
}

@article{das2017keeping,
  title={Keeping the bad guys out: Protecting and vaccinating deep learning with jpeg compression},
  author={Das, Nilaksh and Shanbhogue, Madhuri and Chen, Shang-Tse and Hohman, Fred and Chen, Li and Kounavis, Michael E and Chau, Duen Horng},
  journal={arXiv preprint arXiv:1705.02900},
  year={2017}
}

@article{croce2020robustbench,
  title={Robustbench: a standardized adversarial robustness benchmark},
  author={Croce, Francesco and Andriushchenko, Maksym and Sehwag, Vikash and Debenedetti, Edoardo and Flammarion, Nicolas and Chiang, Mung and Mittal, Prateek and Hein, Matthias},
  journal={arXiv preprint arXiv:2010.09670},
  year={2020}
}

@inproceedings{Croce020a,
  author       = {Francesco Croce and
                  Matthias Hein},
  title        = {Reliable evaluation of adversarial robustness with an ensemble of
                  diverse parameter-free attacks},
  booktitle    = {Proceedings of the 37th International Conference on Machine Learning,
                  {ICML} 2020, 13-18 July 2020, Virtual Event},
  series       = {Proceedings of Machine Learning Research},
  volume       = {119},
  pages        = {2206--2216},
  publisher    = {{PMLR}},
  year         = {2020},
}

@inproceedings{Croce020,
  author       = {Francesco Croce and
                  Matthias Hein},
  title        = {Minimally distorted Adversarial Examples with a Fast Adaptive Boundary
                  Attack},
  booktitle    = {Proceedings of the 37th International Conference on Machine Learning,
                  {ICML} 2020, 13-18 July 2020, Virtual Event},
  series       = {Proceedings of Machine Learning Research},
  volume       = {119},
  pages        = {2196--2205},
  publisher    = {{PMLR}},
  year         = {2020},
}

@misc{guo2024exploringadversarialfrontierquantifying,
      title={Exploring the Adversarial Frontier: Quantifying Robustness via Adversarial Hypervolume}, 
      author={Ping Guo and Cheng Gong and Xi Lin and Zhiyuan Yang and Qingfu Zhang},
      year={2024},
}

@inproceedings{TramerB19,
  author       = {Florian Tram{\`{e}}r and
                  Dan Boneh},
  editor       = {Hanna M. Wallach and
                  Hugo Larochelle and
                  Alina Beygelzimer and
                  Florence d'Alch{\'{e}}{-}Buc and
                  Emily B. Fox and
                  Roman Garnett},
  title        = {Adversarial Training and Robustness for Multiple Perturbations},
  booktitle    = {Advances in Neural Information Processing Systems 32: Annual Conference
                  on Neural Information Processing Systems 2019, NeurIPS 2019, December
                  8-14, 2019, Vancouver, BC, Canada},
  pages        = {5858--5868},
  year         = {2019},
}
